\documentclass[12pt]{article}
\usepackage{amsfonts}
\usepackage{amssymb,letterspace}
\usepackage{graphics,psboxit,amsmath}

\def\hybrid{\topmargin -20pt    \oddsidemargin 0pt
        \headheight 0pt \headsep 0pt
        \textwidth 6.35in       
        \textheight 9.25in       
        \marginparwidth .875in
        \parskip 5pt plus 1pt   \jot = 1.5ex}

\hybrid

\def\baselinestretch{1.2}

\catcode`\@=11

\def\marginnote#1{}
\newcount\hour
\newcount\minute
\newtoks\amorpm
\hour=\time\divide\hour by60
\minute=\time{\multiply\hour by60 \global\advance\minute by-\hour}
\edef\standardtime{{\ifnum\hour<12 \global\amorpm={am}%
        \else\global\amorpm={pm}\advance\hour by-12 \fi
        \ifnum\hour=0 \hour=12 \fi
        \number\hour:\ifnum\minute<10 0\fi\number\minute\the\amorpm}}
\edef\militarytime{\number\hour:\ifnum\minute<10 0\fi\number\minute}

\def\draftlabel#1{{\@bsphack\if@filesw {\let\thepage\relax
   \xdef\@gtempa{\write\@auxout{\string
      \newlabel{#1}{{\@currentlabel}{\thepage}}}}}\@gtempa
   \if@nobreak \ifvmode\nobreak\fi\fi\fi\@esphack}
        \gdef\@eqnlabel{#1}}
\def\@eqnlabel{}
\def\@vacuum{}
\def\draftmarginnote#1{\marginpar{\raggedright\scriptsize\tt#1}}

\def\draft{\oddsidemargin -.5truein
        \def\@oddfoot{\sl preliminary draft \hfil
        \rm\thepage\hfil\sl\today\quad\militarytime}
        \let\@evenfoot\@oddfoot \overfullrule 3pt
        \let\label=\draftlabel
        \let\marginnote=\draftmarginnote
   \def\@eqnnum{(\theequation)\rlap{\kern\marginparsep\tt\@eqnlabel}%
\global\let\@eqnlabel\@vacuum}  }

\def\preprint{\twocolumn\sloppy\flushbottom\parindent 2em
        \leftmargini 2em\leftmarginv .5em\leftmarginvi .5em
        \oddsidemargin -.5in    \evensidemargin -.5in
        \columnsep .4in \footheight 0pt
        \textwidth 10.in        \topmargin  -.4in
        \headheight 12pt \topskip .4in
        \textheight 6.9in \footskip 0pt
        \def\@oddhead{\thepage\hfil\addtocounter{page}{1}\thepage}
        \let\@evenhead\@oddhead \def\@oddfoot{} \def\@evenfoot{} }

\def\numberbysection{\@addtoreset{equation}{section}
        \def\theequation{\thesection.\arabic{equation}}}

\def\underline#1{\relax\ifmmode\@@underline#1\else
        $\@@underline{\hbox{#1}}$\relax\fi}

\def\titlepage{\@restonecolfalse\if@twocolumn\@restonecoltrue\onecolumn
     \else \newpage \fi \thispagestyle{empty}\c@page\z@
        \def\thefootnote{\fnsymbol{footnote}} }

\def\endtitlepage{\if@restonecol\twocolumn \else \newpage \fi
        \def\thefootnote{\arabic{footnote}}
        \setcounter{footnote}{0}}  

\catcode`@=12
\relax

\def\figcap{\section*{Figure Captions\markboth
        {FIGURECAPTIONS}{FIGURECAPTIONS}}\list
        {Figure \arabic{enumi}:\hfill}{\settowidth\labelwidth{Figure
999:}
        \leftmargin\labelwidth
        \advance\leftmargin\labelsep\usecounter{enumi}}}
 \relax
\def\tablecap{\section*{Table Captions\markboth
        {TABLECAPTIONS}{TABLECAPTIONS}}\list
        {Table \arabic{enumi}:\hfill}{\settowidth\labelwidth{Table
999:}
        \leftmargin\labelwidth
        \advance\leftmargin\labelsep\usecounter{enumi}}}
 \relax
\def\reflist{\section*{References\markboth
        {REFLIST}{REFLIST}}\list
        {[\arabic{enumi}]\hfill}{\settowidth\labelwidth{[999]}
        \leftmargin\labelwidth
        \advance\leftmargin\labelsep\usecounter{enumi}}}
 \relax
\makeatletter
\newcounter{pubctr}
\def\publist{\@ifnextchar[{\@publist}{\@@publist}}
\def\@publist[#1]{\list
        {[\arabic{pubctr}]\hfill}{\settowidth\labelwidth{[999]}
        \leftmargin\labelwidth
        \advance\leftmargin\labelsep
        \@nmbrlisttrue\def\@listctr{pubctr}
        \setcounter{pubctr}{#1}\addtocounter{pubctr}{-1}}}
\def\@@publist{\list
        {[\arabic{pubctr}]\hfill}{\settowidth\labelwidth{[999]}
        \leftmargin\labelwidth
        \advance\leftmargin\labelsep
        \@nmbrlisttrue\def\@listctr{pubctr}}}
 \relax
\makeatother
\newskip\humongous \humongous=0pt plus 1000pt minus 1000pt

\newif\ifdtup

\relax

\def\be{\begin{equation}}
\def\ee{\end{equation}}
\def\ba{\begin{eqnarray}}
\def\ea{\end{eqnarray}}

\def\a{\alpha}

\def\th{\theta}

\def\m{\mu}
\def\n{\nu}

\def\l{\lambda}

\def\s{\sigma}

\def\bs{\bigskip}
\def\no{\noindent}

\def\IR{\relax{\rm I\kern-.18em R}}
\def\II{\relax{\rm 1\kern-.35em1}}

\renewcommand{\theequation}{\thesection.\arabic{equation}}
\csname @addtoreset\endcsname{equation}{section}

\def\IR{\relax{\rm I\kern-.18em R}}
\def\inv{^{\raise.15ex\hbox{${\scriptscriptstyle -}$}\kern-.05em 1}}

\begin{document}

\begin{titlepage}
\begin{center}

\hfill NEIP-04-01\\
\vskip -.1 cm
\hfill IFT-UAM/CSIC-04-07\\
\vskip -.1 cm
\hfill hep--th/0403139\\

\vskip .5in

{\LARGE The $SU(3)$ spin chain sigma model and string theory}
\vskip 0.4in

{\bf Rafael Hern\'andez$^1$}\phantom{x} and\phantom{x}
 {\bf Esperanza L\'opez}$^2$ 
\vskip 0.1in

${}^1\!$
Institut de Physique, Universit\'e de Neuch\^atel\\
Breguet 1, CH-2000 Neuch\^atel, Switzerland\\
{\footnotesize{\tt rafael.hernandez@unine.ch}}

\vskip .2in

${}^2\!$
Departamento de F\'{\i}sica Te\'orica C-XI
and Instituto de F\'{\i}sica Te\'orica  C-XVI\\
Universidad Aut\'onoma de Madrid,
Cantoblanco, 28049 Madrid, Spain\\
{\footnotesize{\tt esperanza.lopez@uam.es}}

\end{center}

\vskip .4in

\centerline{\bf Abstract}
\vskip .1in
\no
The ferromagnetic integrable $SU(3)$ spin chain provides the one loop anomalous 
dimension of single trace operators involving the three complex scalars of ${\cal N}=4$ 
supersymmetric Yang-Mills. We construct the non-linear sigma model describing the continuum 
limit of the $SU(3)$ spin chain. We find that this sigma model corresponds to a string 
moving with large angular momentum in the five-sphere in $AdS_5 \times S^5$. The energy 
and spectrum of fluctuations for rotating circular strings with angular momenta along 
three orthogonal directions of the five-sphere is reproduced as a particular case from 
the spin chain sigma model.

\noindent

\vskip .4in
\noindent

\end{titlepage}
\vfill
\eject

\def\baselinestretch{1.2}


\baselineskip 20pt


\section{Introduction}
  
A complete formulation of the AdS/CFT correspondence requires a precise identification 
of states on the string theory side with local gauge invariant operators on the dual 
field theory. But describing this identification in detail is truly involved because it 
implies both understanding the quantization of the string action in $AdS_5 \times S^5$, which 
remains a complicated problem, and obtaining the whole spectrum of ${\cal N}=4$ supersymmetric 
Yang-Mills operators, which is difficult to compute. These difficulties were however 
overcome after the observation that there is a maximally 
supersymmetric plane-wave background for the IIB string \cite{BFHP}, that can be readily 
quantized in the light-cone gauge \cite{Metsaev}. As this plane-wave geometry is obtained 
through a limit of the $AdS_5 \times S^5$ background, the dual description in terms of a 
supersymmetric gauge theory must also involve some sort of equivalent limit. The plane-wave string/gauge 
theory duality is perturbatively reachable from both sides of the correspondence, and provides 
a precise identification relating string states to operators on the field theory side 
carrying a large charge \cite{BMN}. 
  
These gauge theory operators are of the form $\hbox{Tr} (X_1^J \ldots )$,
with $X_1$ one of the ${\cal N}=4$ complex scalars and $J \gg 1$,
and where the dots stand for insertions of few other Yang-Mills fields.
They have a dual realization as small closed strings 
whose center of mass is moving with large angular momentum $J$ along some 
circle of $S^5$ and are localized at the center of $AdS_5$ \cite{GKP}. 
Similarly, operators composed of 
the three ${\cal N}=4$ complex scalars,
$\hbox{Tr} (X_1^{J_1} X_2^{J_2} X_3^{J_3})$, were 
proposed to correspond to semiclassical string solutions with three 
large angular momenta on the five-sphere, $J_1$, $J_2$ and $J_3$ \cite{FT}.
    
In order to prove the duality a comparison must therefore be performed of the spectrum of 
string excitations with the anomalous dimensions of the corresponding field theory operators. However, 
once quantum corrections are taken into account there is operator
mixing, and in order to construct 
generic operators and evaluate their anomalous dimension one should consider a one loop 
mixing for a large number of operators \cite{mixing}. A brilliant insight into this tough 
obstacle came from the observation that the planar one loop anomalous
dimension operator in the scalar sector of 
${\cal N}=4$ supersymmetric Yang-Mills is the hamiltonian of an integrable $SO(6)$ spin chain \cite{MZ}. 
Anomalous dimensions of Yang-Mills operators carrying large charges can then be calculated by solving 
the thermodynamic Bethe ansatz, and successfully compared to the energy
of string solitons \cite{dilatation, spinning}. 
The integrability structure in ${\cal N}=4$ theory is not restricted to the
scalar sector. Indeed, at one loop, it has been proved to extended to all 
conformal operators \cite{Beisert}, leading to an integrable $PSU(2,2|4)$ 
spin chain \cite{psu}. These observations have motivated a wide variety of related 
developments \cite{Aruty}-\cite{Dimov}.

A natural question is then the relation of the integrable spin chain
systems to the string non-linear sigma model. 
This problem was addressed in \cite{Martin}, 
where the continuum limit of the $SU(2)$ Heisenberg spin chain was shown 
to reproduce the action describing strings rotating with large angular 
momentum in an $S^3$ section of $S^5$. 
This identification provides a very powerful tool for analyzing the
integrable structures that arise on both sides of the correspondence, as
well as for improving our understanding of the AdS/CFT correspondence 
itself. A recent paper \cite{Martin2} has shown that the agreement 
between the continuum limit of the spin chain and the string action
holds to two loops in the $SU(2)$ sector. Moreover, 
it has been proved that the thermodynamic limit
of the Bethe equations in this sector precisely coincide with the 
classical results derived from the string sigma model up to two loops
\cite{Kazakov}.

The $SU(2)$ Heisenberg spin chain accounts for the anomalous dimensions of 
${\cal N}=4$ single trace operators of the form $\hbox{Tr} (X_1^{J_1} X_2^{J_2})$
and permutations. The aim of this paper is to extend the work of 
\cite{Martin,Martin2} and study the continuum limit of the $SU(3)$ spin 
chain describing the operators composed of arbitrary combinations of the 
three ${\cal N}=4$ complex scalars. Our analysis will consider only
one loop contributions to the dilatation operator. 
The plan of the paper is the following. In section 2 we will obtain the
non-linear sigma model associated to the continuum limit of the
ferromagnetic $SU(3)$ spin chain.
In section 3 we will show that the sigma model derived from the spin 
chain reproduces the motion of a string with large angular momenta
along three orthogonal directions of the five-sphere. 
A particularly interesting class of solitons in $AdS_5 \times S^5$ is 
that of circular strings \cite{FT,fluctu2}.
In section 4 we will recover their energy and spectrum of fluctuations 
from the spin chain sigma model. We will end with some conclusions and
further directions of research in section 5.


\section{Ferromagnetic Heisenberg chain}

The problem of finding the spectrum of anomalous dimensions for 
${\cal N}=4$ Yang-Mills operators 
$\hbox{Tr} (X_1^{J_1} X_2^{J_2} X_3^{J_3})$, or any of its permutations,
can be mapped to that of solving an
integrable spin chain where at each site sits a fundamental
representation of $SU(3)$. The correspondence between Yang-Mills 
operators and spin chain configurations is
\be
\hbox{Tr} \, (X_{i_1} X_{i_2} X_{i_3} X_{i_4} \dots ) \; \rightarrow \;
 |i_1 \, i_2 \, i_3 \, i_4 \dots \rangle \, ,
\label{tr}
\ee
where $|i_l\rangle$ ($i_l=1,2,3$) expand the fundamental representation
of $SU(3)$ at the $l^{th}$ site. 

The spin chain can be mapped to a discrete sigma model by introducing at 
each site a continuum set of variables. This is reviewed in detail in
many textbooks for the simpler case of the $SU(2)$ spin $s$ 
chain \cite{FAA}. An infinite set of spin states is obtained by applying an 
arbitrary $SU(2)$ rotation to the maximally polarized state at each site.
There is an $U(1)$ subgroup of rotations that just multiplies it 
by a phase, thus mapping this state to itself. Hence two angular variables
are enough to label the resulting set of states. For $s=1/2$, which is 
the relevant case for ${\cal N}=4$ supersymmetric Yang-Mills, we have
\be
|\hat{n} \rangle = \cos \psi \, e^{i \varphi} \, |1\rangle + 
\sin \psi \, e^{-i \varphi} \, |2 \rangle \, ,
\ee
where the states $|1\rangle$ and $|2\rangle$ expand the fundamental 
representation of $SU(2)$.
The central property that makes the overcomplete set $|\hat{n}\rangle$
relevant, is that they provide a resolution of the identity
\be
\int d \m(\hat{n}) \; |\hat{n} \rangle \langle \hat{n} | = \II \ .
\label{idsu2}
\ee
The appropriate measure over the continuum
variables $\psi \in [0,\pi/2]$ and $\varphi \in [0,\pi]$ is
\be
d \mu(\hat{n}) = \frac {4}{\pi} \sin \psi \cos \psi 
\, d \psi \, d \varphi \ .
\ee

A path-integral analysis of the partition function, with the help of 
\eqref{idsu2}, shows the equivalence between the spin chain and a 
discrete sigma model with action
\be 
S= \sum_{l=1}^J \int dt \, \big[\, \omega_l(t) \, - \, {\cal H}_l(t) 
\, \big] \, ,
\label{action}
\ee
with
\begin{eqnarray}
\omega_l(t)  & = & 
{\rm arg} \; \frac {d}{d \rho }\, \langle \hat{n}_l(t) | \hat{n}_l(t+\rho) 
\rangle \big|_{\rho=0} \ ,
\nonumber \\
{\cal H}_l(t) &= & \langle \hat{n}_l(t) , \hat{n}_{l+1}(t) | H_{l,l+1} 
|\hat{n}_l(t) , \hat{n}_{l+1}(t) \rangle \label{Sterms} \, , 
\end{eqnarray}
for a chain governed by nearest neighbor interactions, 
$H=\sum H_{l,l+1}$. 
The continuum limit of this discrete sigma model arises in the
thermodynamic limit of the chain, where the number of 
sites $J \rightarrow \infty$, and only slowly varying spin configurations 
are considered, allowing to replace finite differences between spin 
variables at neighboring sites by derivatives.

In this section we will construct the non-linear sigma model that 
describes the continuum limit of the $SU(3)$ spin chain based on the
fundamental representation. 
We define a set of spin coherent states at each site by applying a 
generic $SU(3)$ rotation to an eigenstate of the Cartan generators.
There is an $SU(2)$ subgroup of rotations that leaves 
it invariant plus an additional $U(1)$ subgroup that just 
multiplies it by a phase. Thus we need four angular variables to label  
the continuum set of states
\be
|\hat{n} \rangle =  
\cos \theta \, \cos \psi \, e^{i \varphi} \, |1 \rangle + 
\cos \theta \, \sin \psi \, e^{-i \varphi} \, |2 \rangle + 
\sin \theta \, e^{i \phi} \, |3 \rangle\ \ .
\label{cont}
\ee
The variables $\phi$ and $\varphi$ are related to rotations generated
by the two Cartan elements of $SU(3)$, while $\theta$ and $\psi$
correspond to
rotations generated by two non-Cartan elements acting on different subspaces.
The states $|\hat{n}\rangle$ form an overcomplete basis, with the resolution 
of identity
\be
\int d \m(\hat{n}) \; |\hat{n} \rangle \langle \hat{n} | = \II \ ,
\label{id}
\ee
where now the measure is
\be
d \mu(\hat{n}) = \frac {12}{\pi^2} \,
\sin \theta \cos^3 \theta \sin \psi \cos \psi 
\, d \theta \, d \psi \, d \phi \, d \varphi \ .
\label{measure}
\ee
Ranges of these variables are
$\theta, \psi \in [0,\pi/2]$ and $\phi+\varphi, \phi-\varphi \in [0,2\pi]$.

We analyze next the hamiltonian of the $SU(3)$ spin chain. The
hamiltonian of the $SO(6)$ closed quantum spin chain with nearest
neighbor interactions was recovered in \cite{MZ} from the planar one loop 
anomalous dimension operator 
in the scalar sector of ${\cal N}=4$ Yang-Mills
\be
H = \frac {\l}{16\pi^2} \sum_{l=1}^J (K_{l,l+1}+2-2P_{l,l+1}) \ .
\ee
The trace operator $K_{l,l+1}$ is zero when acting on the $SU(2)$ or
$SU(3)$ sectors, namely on operators composed out of the ${\cal N}=4$
complex scalars without inclusion of their complex conjugates. Thus 
the hamiltonian in this sectors reduces to
\be
H= \frac {\lambda}{8 \pi^2} \sum_{l=1}^J (1-P_{l,l+1}) \, ,
\label{heise}
\ee
where $P_{l,l+1}$ is the permutation operator between the sites $l$ and
$l+1$. For the $SU(2)$ case
\eqref{heise} is just the hamiltonian for the ferromagnetic XXX
Heisenberg spin chain.

The symmetry group of the chain can be extended from $SU(3)$ to $U(3)$
by adding a trivial $U(1)$. This $U(1)$ acts with the same rotation on 
each site of the chain. Its associated charge is the total number of sites,
$J$. A basis of $U(3)$ generators acting on each site is
\be
(T^{ab})_{ij}= \delta^a_i \delta^b_j \ ,
\ee
where $a,b,i,j=1,2,3$. In terms of this basis the permutation operator
is given by
\be
P_{l,l+1}= T^{ab}_l T^{ba}_{l+1} \ .
\ee
Let us consider two states, 
$|\alpha \rangle = \sum \alpha_i \, |i \rangle$ and 
$|\beta \rangle = \sum \beta_i \, |i \rangle$, located respectively at 
sites $l$ and $l+1$ of the chain and compute the matrix element
\be
\langle \alpha, \beta | P_{l,l+1} |\alpha, \beta \rangle
= \langle \alpha |T^{ab}_l|\alpha \rangle \; \langle \beta 
|T^{ba}_{l+1}|\beta \rangle = 
|\langle \alpha |\beta \rangle|^2 \ .
\ee

Therefore, in order to derive the sigma model associated to the chain, 
all what we need to analyze is the scalar product for two coherent
states of the form \eqref{cont}
\begin{eqnarray}
\langle \hat{n}'|\hat{n} \rangle & = & 
\sin \theta \sin \theta' \cos (\phi-\phi')  + 
\cos \theta \cos \theta' \cos (\psi-\psi') 
\cos (\varphi-\varphi')+ \nonumber \\
&+ & i \left[ \sin \theta \sin \theta'\sin (\phi-\phi') 
+ \cos \theta \cos \theta' \cos (\psi+\psi') 
\sin (\varphi-\varphi')\right] \ .
\end{eqnarray}
From \eqref{action} and \eqref{Sterms}, in the long wavelength limit of
the chain, where $|\hat{n} \rangle-|\hat{n}'\rangle=
|\delta \hat{n} \rangle$, we obtain
\begin{eqnarray}
S & = & \frac {J}{ 2 \pi} \int d\s \, dt \big( \sin^2 \theta \; 
\dot \phi + \cos^2 \theta \, \cos (2 \psi) \;
\dot \varphi \big)- \frac {\l}{4 \pi J} \int d\s \, dt 
\big[ \theta'^2 \nonumber \\ 
& + & \cos^2 \theta \, \left( \psi'^2 + \sin^2 (2 \psi) \, \varphi'^2  \right)
+ \sin^2 \theta \, \cos^2 \theta \, \left( \phi'- \cos (2 \psi) \, \varphi' 
\right)^2  \big]\ ,
\label{su3action}
\end{eqnarray}
where the dot and prime stand for derivatives with respect to $t$ and
$\s$, respectively. The 
result for the $SU(2)$ sigma model \cite{Martin} can be recovered by setting
$\theta=0$. The kinetic piece in the action is a Wess-Zumino term.
The trace in the Yang-Mills operators implies that we should consider
only periodic spin configurations invariant under the shift operator
\cite{MZ}. Namely, we require $|\hat{n}(2\pi,t)\rangle=
\pm |\hat{n}(0,t)\rangle$ \footnote{The vectors 
$\pm|\hat{n}\rangle$ represent the same physical state. The choice of
variables for the coherent states \eqref{cont} does not fix this sign
ambiguity, which corresponds to shift 
$\varphi,\phi \rightarrow \varphi \pm\pi, \phi
\pm\pi$.}
together with \cite{Kazakov,Martin}
\be
\int d\sigma \, \cos^2\theta \, \big(\phi'- \cos(2\psi) \varphi' \big) =
2 \pi s \, ,
\label{P}
\ee
with $s$ integer.

The action \eqref{su3action} is invariant under constant shifts in 
$\phi$ and $\varphi$. The corresponding conserved angular momenta are
\be
P_{\phi} = \frac {J}{2 \pi}\int d\s \sin^2 \th \ , \:\:\:\:\:\: P_{\varphi} = 
\frac {J}{2 \pi} \int d\s \cos^2 \th \cos (2\psi) \ ,
\label{Ps}
\ee
and the hamiltonian is
\be 
H = \frac {\l}{4 \pi J} \int \! d\s \, \big[ \theta'^2 + \cos^2 \theta \, 
\left( \psi'^2 + \sin^2 (2 \psi) \, \varphi'^2  \right)
+ \frac {1}{4} \sin^2 (2\theta) \, \left( \phi'- \cos (2 \psi) \, \varphi' 
\right)^2  \big] \ .
\ee  

 
\section{String sigma model}

We will now describe how the non-linear sigma model for a 
string rotating in $S^5$ becomes the ferromagnetic sigma model
for the $SU(3)$ spin chain, after some adequate large angular momentum 
limit is taken. Let us then consider 
the propagation of the dual string on $S^5$. The associated metric, 
including the decoupled time coordinate $t$,
\be
ds^2= -dt^2 + d \theta ^2 + \sin^2 \theta \, d \phi_3^2 +
\cos^2 \theta \left( d \psi^2 + \cos^2 \psi \, d \phi_1^2 + 
\sin^2 \psi \, d \phi_2^2 \right) \ ,
\label{metric}
\ee
can be more conveniently written if we define  
\be
\phi_1= \alpha + \varphi \; , \hspace{.5cm} \phi_2 = \alpha - \varphi 
\; , \hspace{.5cm} \phi_3= \alpha+\phi \ .
\label{ch}
\ee
The metric then becomes
\begin{eqnarray}
ds^2 \!\! & = & \!\! -dt^2 + d \theta ^2 + \sin^2 \theta \, d \phi^2 +
2 \sin^2 \theta \, d \phi \, d \alpha + d \alpha^2 \nonumber \\
& + & \!\! \cos^2 \theta \left(d \psi^2 + 2 \cos (2\psi)\, d \varphi\,  d \alpha
+d \varphi^2  \right) \ .
\end{eqnarray}
Introducing one more change of variables, $\alpha \rightarrow \alpha +t$, we obtain
\begin{eqnarray}
ds^2 \!\! & = & \!\! d \theta ^2 + \sin^2 \theta \, d \phi^2 +
2 \sin^2 \theta \, d \phi \left(dt+ d \alpha\right) + 2 dt \, d \alpha + d\a^2 \nonumber \\ 
& + & \!\! \cos^2 \theta \left(d \psi^2 + 2 \cos (2\psi)\, d \varphi 
\left( dt + d \alpha \right) + d \varphi^2 \right) \ .
\label{s5}
\end{eqnarray}
  
We compute next the Polyakov action in this background. We will  
choose coordinates so that $t=\kappa \tau$. An interesting
limit to consider is $\kappa \rightarrow \infty$, while 
$\kappa \, \partial_\tau X^\m$ (with $X^\mu \neq t$) is kept fixed 
\cite{Martin}. This limit corresponds to a string whose motion is mainly 
absorbed by the shift in the coordinate $\alpha$, and gives rise to 
large spins: $\partial_\tau \phi_i = \kappa + 
{\cal O}\big(\frac {1}{\kappa})$. In this limit the action 
describing a string in the background (\ref{s5}) simplifies to 
\ba
S & = & \frac {R^2}{4\pi \a'} \int d\s \, d\tau \big[ 2 \kappa \,
\partial_\tau \a 
+ 2 \kappa \cos^2 \th \cos (2\psi) \, \partial_\tau \varphi + 2 \kappa \sin^2
\th \, \partial_\tau \phi - \a'^2 - \th'^2 \nonumber \\ 
& - & \cos^2 \th \, \varphi'^2 - \sin^2 \th \, \phi'^2 - \cos^2 \th \, \psi'^2 
- 2 \sin^2 \th \, \a' \phi' - 2 \cos^2 \th \cos (2 \psi) \, \a' \varphi' \big] \ , 
\label{Slimit}
\ea
and the Virasoro constraints, $G_{\m\n} \partial_{\tau}X^{\m} \partial_{\s} X^{\n}=0$ and 
$G_{\m\n} \partial_{\tau} X^{\m} \partial_{\tau} X^{\n} + 
G_{\m\n} \partial_{\s} X^{\m} \partial_{\s} X^{\n}=0$, reduce, respectively, to
\be
2 \kappa \a' + 2 \kappa \cos^2 \th \cos (2 \psi) \, \varphi' + \, 2 \kappa \sin^2 \th \, \phi' = 0 \ , 
\label{v1} 
\ee
and 
\ba
&& 2 \kappa \, \partial_\tau \a + 2 \kappa \cos^2 \th \cos (2\psi) \, 
\partial_\tau \varphi + 2 \kappa \sin^2 \th \, \partial_\tau \phi 
+ \a'^2 + \th'^2 + \cos^2 \th \, \varphi'^2 \nonumber \\ 
&& + \sin^2 \th \, \phi'^2 + \cos^2 \th \, \psi'^2 + 2 \sin^2 \th \, \a' \phi' 
+ 2 \cos^2 \th \cos (2 \psi) \, \a' \varphi' = 0 \ . \label{v2}
\ea
If we insert (\ref{v1}) in (\ref{Slimit}), and change variables to
$t= \kappa \tau$, we get
\ba
S & = & \frac {R^2 \kappa}{2\pi \a'} \int d\s \, dt \big[ \dot{\a} 
+ \cos^2 \th \cos (2\psi) \, \dot{\varphi} + \sin^2 \th \, \dot{\phi}
\big] - \frac {R^2}{4 \pi \alpha'\kappa} \int d\s \, dt \big[
\th'^2 \nonumber \\
& + & \cos^2 \th ( \psi'^2 + \sin^2 (2\psi) \, \varphi'^2 ) + \sin^2 
\th \cos^2 \th (\phi' - \cos (2 \psi) \, \varphi')^2 \big] \ . 
\ea
The momentum associated to $\alpha$ is $P_\alpha=\kappa \, R^2/\alpha'$.
According to the change of coordinates \eqref{ch}, $P_\alpha$ should
be identified with the total angular momentum, $J$. Therefore this 
action coincides precisely with the $SU(3)$ spin chain sigma model
action (\ref{su3action}), once we use the gravity/gauge theory relation
$R^2=\a' \sqrt{\l}$. 
Notice that a trivial variable playing the same
role as $\alpha$ could have been introduced also in the spin chain 
sigma model, where it would correspond to an irrelevant global phase 
multiplying the 
state $|\hat{n}\rangle$. The measure on the overcomplete set
$\{|\hat{n}\rangle\}$ introduced in the previous section \eqref{measure} is
just (proportional to) the volume element of the five-sphere.


\section{Circular strings}

In this section we will concentrate on a particular family of classical
string solutions, that of circular strings rotating on $S^5$ and
located at the center of $AdS_5$
\cite{FT,fluctu2}. Using the spin chain sigma model we will reobtain 
their energy and spectrum of fluctuations at first
order in the effective coupling constant $\frac {\l}{J^2}$, 
verifying that the limit we have taken in the string sigma model
correctly reproduces the known results for large angular momentum. 
It would be interesting to extend this 
analysis to other classes of solutions (see \cite{Tseytlin} for a complete 
account of semiclassical string solutions).
  
The equations of motion for the spin chain sigma model 
(\ref{su3action}) 
are
\ba
&& \!\!\!\!\!\! \sin (2\psi) \Big[ \cos^2 \th \, \dot{\psi} + 
\frac {\l}{4 J^2}
\sin^2(2 \th)\big(\phi'-\cos(2\psi) \varphi'\big) \psi'\Big] + 
\frac {\l}{2J^2} \partial_{\s} \big[ \cos^2 \th \sin^2 (2\psi) 
\varphi' \big] = 0 \ , \nonumber \\
&& \!\!\!\!\!\! \sin (2\th) \, \dot{\th} - \frac {\l}{4J^2} 
\partial_{\s} \big[ \sin^2 (2\th) (\phi'- \cos(2\psi) \varphi' ) 
\big] = 0 \ , \label{eom} \\
&& \!\!\!\!\!\! \cos^2 \th \sin (2\psi) \Big( \dot{\varphi} + 
\frac {\l}{J^2} 
\big[ \cos^2 \th \, \cos (2\psi) \varphi'^2 + \sin^2 \th \, \varphi'
\phi') \big] \Big) - \frac {\l}{2J^2} \partial_{\s} 
\big[ \cos^2 \th \psi'\big] = 0 \ , \nonumber 
\ea
\ba
&& \!\!\!\!\!\! \frac {\l}{J^2} \th'' \! + \sin (2\th) \Big( \! 
\dot{\phi}-\cos (2\psi) \dot{\varphi}  + \frac {\l}{2J^2} 
\big[ \psi'^2 \! + \sin^2 (2\psi) \varphi'^2 \! 
- \cos(2\th) \big( \phi' \! - \cos(2\psi) \varphi' \big)^2 \big] \Big)  =
0 \nonumber \ .
\ea
  
Circular string solutions correspond to the ansatz $\th=\th_0$ and 
$\psi=\psi_0$ constant, and $\partial_{\s} \varphi = m$, 
$\partial_{\s} \phi = n$, where $m$ and $n$ are both integers
or half-integers fulfilling \eqref{P}.
This ansatz solves the previous equations with
\ba
\dot{\varphi} &=& - \frac {\l}{J^2} m \Big[\, m \Big(\frac {J_1}{J}-
\frac{J_2}{J}\Big)+ n \frac {J_3}{J} \Big] \ ,
\nonumber \\
\dot{\phi}&=& \frac {\l}{2 J^2} \Big( n^2 -m^2 - 2n \Big[\, m \Big(
\frac {J_1}{J}-\frac {J_2}{J}\Big)+
n \,\frac {J_3}{J} \Big] \Big) \ ,
\label{ws}
\ea
where we have made use of the relations
\be
\frac {J_1-J_2}{J_1 + J_2}= \cos (2\psi_0) \ , \hspace{1cm}
\frac {J_3}{J}= \sin^2 \th_0 \ ,
\ee
derived from the conserved momenta \eqref{Ps}, $P_\varphi=J_1-J_2$ and 
$P_\phi=J_3$. We obtain the following expression for the energy 
\be
E = \frac {\lambda}{2J}  \frac {1}{J^2} \big[(2m)^2 J_1J_2 + (n-m)^2 J_1 J_3 + (n+m)^2 J_2 J_3 \big] \ .
\label{e}
\ee
Equations \eqref{ws} and \eqref{e} reproduce the angular velocities 
and energy of a circular string rotating along three orthogonal
directions of $S^5$ \cite{fluctu2}, provided we identify
\be
2m = m_1-m_2 \ , \:\:\:\: n-m = m_3-m_1 \ , \:\:\:\: n+m = m_3-m_2 \ ,
\label{iden}
\ee
If we further equal $s=m_3$, \eqref{P} reproduces the well-known 
constraint of circular strings $\sum_{i=1}^3 m_i J_i=0$.
Notice that \eqref{iden} is consistent with \eqref{ch}; an analogous 
relation applies for the angular velocities.

The study of the spectrum of quadratic fluctuations around these
solutions allows to determine their stability and to derive the one loop
sigma model corrections to the energy \cite{fluctu1,fluctu2}. 
In the limit of large angular momenta the spectrum of excitations
contains modes with low frequencies $\omega \sim \frac {\l}{J^2}$, and
modes with frequencies $\omega \sim 1$. We will now
check that the low frequency fluctuations can also be derived from the
spin chain sigma model.

Let us first consider a two-spin circular string. They are obtained by 
setting in the previous ansatz $\th_0=0$, which implies $J_3=0$. 
With this restriction, equations \eqref{eom} reduce to those of the
$SU(2)$ spin chain sigma model \cite{Martin}. Quadratic fluctuations around
circular two-spin solutions are governed by the equations
\ba
&& \dot{\psi} + \frac {\l}{2 J^2} \big( 4 m \cos (2\psi_0) \, \psi'+
\sin (2\psi_0) \, \varphi'' \big)=0 \ , \nonumber \\
&& \sin^2(2\psi_0) \dot{\varphi} - \frac {\l}{2 J^2} \big( \psi''+
4 m^2 \sin^2 (2\psi_0) \, \psi - 2m \sin (4\psi_0) \, \varphi' \big) =0 \ . 
\ea
We easily obtain the spectrum
\be
\omega = \frac {\l}{2J^2} r \Big[ -4m \cos (2 \psi_0)
\pm \sqrt{r^2 - 4 m^2 \sin^2 (2 \psi_0)} \: \Big] \ , 
\ee
with $r$ integer, which reproduces the results in \cite{fluctu2}.
For the fluctuations, \eqref{P} translates into the zero momentum constraint
$\sum_r N_r \,r=0$, with $N_r$ the number of modes of momentum
$r$.

The analysis of fluctuations around generic three-spin circular solutions
is straightforward but tedious. Here we will only consider the simplified
case when $J_1=J_2$ and $n=0$. In terms of the string sigma model
variables this translates into $J_1=J_2$ and $m_3=0$, which was analyzed 
in \cite{fluctu1}. The linearized equations of
motion around this solution are
\ba
&& \frac {\l}{J^2} \, \th'' + \sin(2\th_0) \, \dot{\phi} + 
\frac {\l}{J^2}\, m  \sin (2\th_0) \, \varphi' =0 \ , \nonumber \\
&& \dot{\th} - \frac {\l}{4 J^2} \sin(2\th_0) \, 
\big(\phi''+2m \psi'\big) =0 \ , \\
&& \dot{\psi} - \frac {\l}{2 J^2} \, \big( 2 m \tan \th_0 \, 
\th'-\varphi''\big)=0 \ , \nonumber \\
&& \dot{\varphi} - \frac {\l}{J^2} \, \big( 2m^2 \cos^2 \th_0 \, \psi -
m \sin^2 \th_0 \, \phi'\big) - \frac {\l}{2J^2} \, \psi''=0 \ . \nonumber
\ea
We obtain the following spectrum of fluctuations
\ba
&& \!\!\!\!\!\!\!\!\!\!\!\!\!\!\!\!\!\! 
\omega^2 \! = \! \Big(\frac {\l}{2 J^2} \Big)^{\!2} r^2 \! \left[r^2 \! + 2 m^2
\big(3 \sin^2 \! \th_0 \! -1 \! \big) \pm 2m \sqrt{\big(3 \sin^2 \! \th_0 \! 
- \! 1 \big)^2 m^2 \! + \! 4 \sin^2 \th_0 (r^2 \! - \! m^2)} \, \right] \: ,
\ea
in agreement with \cite{fluctu1}.

The stability properties of the circular strings are determined by
the low frequency fluctuations. This is not the case for the one loop 
corrections to their energy. Although $\Delta E/E \sim 1/J$, the
contributions to $\Delta E$ are not saturated by the modes we have 
obtained above \cite{fluctu1,fluctu2}.
It would be interesting to see whether the one loop correction to the 
classical energy of circular strings can be also derived in the
framework of the spin chain sigma model.


\section{Conclusions}

The motivation of this note has been to explore the relation of the
integrable $SU(3)$ spin chain with the string non-linear sigma model. We 
have in particular precisely identified the ferromagnetic sigma model 
describing the continuum limit 
of the $SU(3)$ chain with the sigma model for a string in $S^5$. In 
order to understand and clarify this equivalence we have 
also seen how classical solutions to the spin chain sigma model 
correspond to string configurations in the dual $AdS_5 \times S^5$ background. 
 
In a recent paper \cite{Martin2}, it has been shown that the
identification between the continuum limit of the spin chain and the 
string action extends to two loops in the $SU(2)$ sector.
The derivation of the spin chain sigma model, which to leading order 
requires considering only long wavelength spin configurations, beyond one 
loop involves quantum corrections from short wavelength configurations. 
Again for the $SU(2)$ sector, the direct correspondence between solutions of 
thermodynamic Bethe ansatz equations and classical solutions of the string 
sigma model to two loop order has been proved in \cite{Kazakov}, 
and shows the equivalence between the integrable
structures arising in the gauge and string theory sides
(see also \cite{AS03,Engquist,AS}).
  
It would be extremely interesting to extend the results in
\cite{Martin2,Kazakov} to the $SU(3)$ sector. However, there is an 
important difference between the $SU(2)$ and $SU(3)$ scalar
sectors of ${\cal N}=4$ Yang-Mills. $SU(2)$ operators
mix only among themselves at all orders in perturbation theory,
maintaining the notion of the length of the chain as a well
defined concept. However, the $SU(3)$ scalar sector 
is only closed at one loop. At higher orders transitions are possible
between the configurations $X_1 X_2 X_3$ and $\Psi_1 \Psi_2$, 
where $\Psi_i$ denote the two complex fermions of ${\cal N}=4$ which are
$SU(3)$ singlets. In order to obtain a closed sector, $SU(3)$ 
has to be extended to $SU(2|3)$, whose fundamental representation can be
associated with both $X_i$ and $\Psi_j$ fields \cite{Beisert}. 
The spin chain corresponding to this sector has the striking property
that the number of sites becomes a dynamical variable. In spite of that, 
arguments were presented in \cite{su23} in favor of its integrability.
The study of this dynamical spin chain is important for
a better understanding of the gauge theory/gravity correspondence.

One should at any rate stress that there is still no clear proof that 
integrability is valid in the planar limit of ${\cal
N}=4$ Yang-Mills beyond two loops \cite{dilatation,SS,Kazakov,AS,Martin2}, 
although we know that classical 
integrability of the string sigma model in $AdS_5 \times S^5$ 
holds to all orders \cite{Spenta,Polchinski,Vallilo}. The analysis of the
continuum limit of the spin chain provides a very promising tool
for addressing this problem, whose expected outcome is that the sigma
model resulting from the spin chain including all higher loop
corrections is the classical action of the string in $AdS_5 \times S^5$
\cite{Martin,Martin2}.


\bs\bs

\centerline {\bf Acknowledgments}

It is a pleasure to thank E. \'Alvarez, C. G\'omez, K. Landsteiner,
G. Sierra and M. Staudacher for useful 
discussions. R.H. acknowledges the financial support provided through
the European
Community's Human Potential Programme under contract HPRN-CT-2000-00131
``Quantum Structure of Space-time'', the Swiss Office for Education 
and Science and the Swiss National Science Foundation. The work of E.L. 
was supported in part by the Spanish DGI of the MCYT under
contract FPA2003-04597.


\end{document}